\documentclass[11pt]{article}
\usepackage{epsfig}
\usepackage{amssymb}
\usepackage{color}
\usepackage{epsf,amsfonts }
\usepackage{amssymb, amsmath, amsthm}
\usepackage[colorlinks=true, pdfstartview=FitV, linkcolor=blue, citecolor=blue, urlcolor=blue]{hyperref}

\newcommand\encadremath[1]{\vbox{\hrule\hbox{\vrule\kern8pt
\vbox{\kern8pt \hbox{$\displaystyle #1$}\kern8pt}
\kern8pt\vrule}\hrule}}
\def\enca#1{\vbox{\hrule\hbox{
\vrule\kern8pt\vbox{\kern8pt \hbox{$\displaystyle #1$}
\kern8pt} \kern8pt\vrule}\hrule}}

\def\ds{\displaystyle}

\def \d {\mathrm d}

\def \restr{\mathrm {resTr}}
\def\res{\mathop{\mathrm {res}}\limits_}

\newtheorem{theorem}{Theorem}[section]

\newtheorem{examps}{Examples}[section]

\newtheorem{lemma}{Lemma}[section]
\newtheorem{remark}{Remark}[section]

\newtheorem{remarks}[remark]{Remarks}
\newtheorem{proposition}{Proposition}[section] 
\newtheorem{definition}{Definition}[section]
\def\H{{\cal H}}

\def\le{\left}
\def\CC{\mathcal C}
\def\bx{\begin{example}}
\def\ex{\end{example}}
\def\bxs{\begin{examps}. \rm\begin{enumerate}}
\def\exs{\end{enumerate}\end{examps}}
\def\bd{\begin{definition}}
\def\ed{\end{definition}}
\def\bt{\begin{theorem}}
\def\et{\end{theorem}}
\def\bp{\begin{proposition}\rm}
\def\ep{\end{proposition}}
\def\bc{\begin{corollary}}
\def\ec{\end{corollary}}
\def\bl{\begin{lemma}\em}
\def\el{\end{lemma}}
\def\be{\begin{equation}}
\def\ee{\end{equation}}
\def\bea{\begin{eqnarray}}
\def\eea{\end{eqnarray}}
\def\beq{\begin{equation}}
\def\eeq{\end{equation}}
\newcommand\Res{\mathop{{\rm Res}}}

\def\GC{{\check{\Gamma}}}
\def\G{\Gamma}
\def\Ib{{\mathbf I}}
\def\Hb{{\mathbf H}}

\def\HbT{\hat{\mathbf H}}
\def\eb{{\mathbf e}}
\def\kappab{{ \mathbf \kappa}}
\def\PsiN{\ds{\mathop{\Psi}_N}}

\renewcommand{\theequation}{\arabic{section}.\arabic{equation}}
\makeatletter
\@addtoreset{equation}{section}

\def\bx{\begin{example}}
\def\ex{\end{example}}
\def\bxs{\begin{examps}. \rm\begin{enumerate}}
\def\exs{\end{enumerate}\end{examps}}
\def\bd{\begin{definition}}
\def\ed{\end{definition}}
\def\bt{\begin{theorem}}
\def\et{\end{theorem}}
\def\bp{\begin{proposition}\rm}
\def\ep{\end{proposition}}
\def\bc{\begin{corollary}}
\def\ec{\end{corollary}}
\def\bl{\begin{lemma}\em}
\def\el{\end{lemma}}
\def\be{\begin{equation}}
\def\ee{\end{equation}}
\def\br{\begin{remark}\rm\small}
\def\er{\end{remark}}
\def\brs{\begin{remarks}.\\ \rm\
\begin{enumerate}}
\def\ers{\end{enumerate}\end{remarks}}
\def\bea{\begin{eqnarray}}
\def\eea{\end{eqnarray}}

\def \pa{\partial}

\def\Tr{\mathrm {Tr}}
\def\tr{\mathrm {tr}}
\def\det{\mathrm {det}}
\def\ln{\mathrm {ln}}

\def\res{\mathop{\mathrm {res}}\limits_}
\def\ri{\right}
\def\ds{\displaystyle}

\def\&{&{\hskip -20pt}}

\def\s{{\sigma}}

\def\GH{{\hat{\Gamma}}}

\def\Cbb{{\mathbb C}}
\def\Nbb{{\mathbb N}}

\begin{document}

\baselineskip 16pt 
\begin{center}
\begin{Large}\fontfamily{cmss}
\fontsize{17pt}{27pt}
\selectfont
\medskip
\textbf{The partition function of  the two-matrix model as an isomonodromic tau-function}
\end{Large}\\
\bigskip
\begin{large} 
 {M. Bertola}$^{\ddagger,\sharp}$\footnote{bertola@crm.umontreal.ca},  {O. Marchal}$^{\dagger, \sharp}$\footnote{olivier.marchal@polytechnique.org}
\end{large}
\\
\bigskip
\begin{small}
$^{\dagger}$ {\em Institut de Physique Th\'eorique,
CEA, IPhT, F-91191 Gif-sur-Yvette, France
CNRS, URA 2306, F-91191 Gif-sur-Yvette, France

$^\sharp$
Centre de recherches math\'ematiques,
Universit\'e de Montr\'eal\\ C.~P.~6128, succ. centre ville, Montr\'eal,
Qu\'ebec, Canada H3C 3J7} \\
\smallskip
$^{\ddagger}$ {\em Department of Mathematics and
Statistics, Concordia University\\ 1455 de Maisonneuve W., Montr\'eal, Qu\'ebec,
Canada H3G 1M8} \\ 
\end{small}
\end{center}
\bigskip
\smallskip
\bigskip
\begin{center}{\bf Abstract}
\end{center}
\smallskip
We consider the Itzykson-Zuber-Eynard-Mehta two-matrix model and prove that the partition function is an isomonodromic tau function in a sense that generalizes Jimbo-Miwa-Ueno's  \cite{JMI}. 
In order to achieve the generalization we need to define a notion of tau-function for isomonodromic systems where the $ad$--regularity of the leading coefficient is not a necessary requirement.

\bigskip
\bigskip
\tableofcontents
\section{Introduction}

Random matrices models have been studied for years and have generated important results in many fields of both theoretical physics and mathematics. 
The two-matrix model 
\be
\d\mu(M_1,M_2) = {\rm e}^{-\Tr (V_1(M_1) + V_2(M_2) -M_1M_2)} \d M_1\d M_2\label{101}
\ee
was used to model $2D$ quantum gravity \cite{DKK} and was investigated from a more mathematical point of view in \cite{MS, EM, BEH1,BEH2,BEH3,BEH4, BE}; the {\em partition function} of the model
\be
\mathcal Z_N(V_1,V_2) = \int\int \d\mu(M_1,M_2)
\ee
has important properties in the large $N$--limit for the enumeration of discrete maps on surfaces \cite{DfGZ} of arbitrary genus and it is also known to be a tau-function for the $2$--Toda hierarchy.
In the case of  the Witten conjecture, proved by Kontsevich \cite{Kontsevich} with the use of matrix integrals not too dissimilar from the above one,  the enumerative properties of the tau function imply some nonlinear (hierarchy of) PDEs (the  KdV hierarchy for the mentioned example). 
On a similar level,  one expects some hierarchy of PDEs  for the case of the two-matrix model and possibly some  Painlev\'e\ property (namely the absence of movable essential singularities). 
The Painlev\'e\ property is characteristic of tau-functions  for isomonodromic families of ODEs that depend on parameters; hence a way of establishing such property for the partition function $\mathcal Z_N$ is that of identifying it with an instance of isomonodromic tau function \cite{JMI, JMII}. 

 This is precisely the  purpose of this article; we capitalize on previous work that showed how to relate the matrix model to certain biorthogonal polynomials \cite{MS, EM} and how these appear in a natural fashion as the solution of certain isomonodromic family \cite{BEH} .
 
 The paper extends to the case of the two matrix model the work contained in \cite{BEH, BEH4, BertoGekhtman}; it uses, however, a different approach, closer to the recent \cite{BE_mom}. 
 
 In \cite{BEH, BEH4, BertoGekhtman, ITW} the partition function of the one--matrix model (and certain shifted T\"oplitz determinants) were identified as isomonodromic tau functions by using {\em spectral residue formul\ae} in terms of the spectral curve of the differential equation. 
 Such spectral curve has interesting properties inasmuch as --in the one-matrix case-- the spectral invariants can be related to the expectation values of the matrix model. Recently the spectral curve of the two matrix model \cite{BE} has been written explicitly in terms of expectation values of the two--matrix model and hence one could use their result and follow a similar path for the proof as the one followed in \cite{BEH4}. 
Whichever one of the two approaches one chooses,  a main obstacle  is that the definition of isomonodromic tau function \cite{JMI, JMII} relies on a genericity assumption for the ODE which fails in the case at hand, thus requiring a generalization in the definition.
 
 According to this logic, one of the purposes of this paper  
  is to extend the notion of tau-function introduced by Jimbo-Miwa-Ueno's  \cite{JMI}, to the two-matrix Itzykson-Zuber model. This task is accomplished in a rather general framework in Sec. 
\ref{taudef}. 

We then show that the partition function has a very precise relationship with the tau-function so introduced, allowing us to (essentially) identify it as an isomonodromic tau function (Thm. \ref{main}).

\section{A Riemann Hilbert formulation of the two-matrix model}
\label{2mmRHP}
According to the seminal work \cite{MS, EM} and 
following the notations and definitions introduced in \cite{BEH1, BEH2}, we consider paired sequences of monic polynomials $\{\pi_m(x), \s_m(y)\}_{m=0\dots \infty}$
 $ (m=\deg{\pi_m} =\deg{\s_m}$), that are biorthogonal in the sense that
 \be
 \int \!\!\! \int _\varkappa dx dy \pi_m(x)\s_n(y) e^{-V_1(x) -V_2(y) +xy} =  h_m \delta_{mn}, \quad h_m\neq 0.
 \ee
The functions $V_1(x), V_2(y)$ appearing here are referred to as  {\it potentials}, terminology drawn from random matrix theory, in which such quantities play a fundamental role.
 
  Henceforth, the second potential $V_2(y)$ will be chosen as a polynomial of degree $d_2+1$
  \be
  V_2(y) = \sum_{j=1}^{d_2+1} \frac{v_j}{j} y^j , \quad v_{d_2+1}  \ \ne 0
  \ee
  For the purposes of most of the considerations to follow, 
  the first potential $V_1(x)$ may have very general analyticity properties as long as the manipulations make sense, but for definiteness and clarity we choose it to be polynomial as well.
  
 The symbol $\int\int_\varkappa$ stands for any linear combination of integrals of the form
 \be
\int\!\!\! \int _\varkappa  dx dy  := \sum_{j}\sum_k \varkappa_{jk}\int_{\G_j} dx \int_{\GH_k} dy ,\qquad \varkappa_{ij} \in \mathbb C
\ee
 where  the  contours $\{\GH_k\}_{k=1\dots d_2}$ will be chosen as follows. In the $y$--plane, define  $d_2+1$ ``wedge sectors'' $\{ \hat{S}_k\}_{k=0\dots d_2}$  such that $\hat{S}_k$ is bounded by the pairs of rays: $r_k:= \{y \vert  \arg{y}= \theta + \frac{2k\pi}{d_2+1}\}$ and $r_{k-1}:= \{y \vert \arg{y}= \theta +\frac{2(k-1)\pi}{d_2 +1}\}$, where $\theta:= \arg{v_{d_2+1}}$. Then $\GH_k$ is any smooth oriented contour  within the sector $\hat{S}_k$ starting from $\infty$ asymptotic to the ray $r_k$ (or any ray within the sector that is at an angle $< \frac{\pi}{2(2d_2+1)}$ to it, which is equivalent for purposes of integration), and returning to $\infty$ asymptotically along $r_{k-1}$ (or at an angle  $ < \frac{\pi}{2(2d_2+1)}$ to it). These will  be referred to as the ``wedge contours''. We also define a set of smooth oriented contours $\{\GC_k\}_{k=1, \dots d_2}$,  that have intersection matrix $\GC_j \cap \GH_k= \delta_{jk}$ with the $\GH_k$'s, 
  such that $\GC_k$ starts from $\infty$ in sector $\hat{S}_0$, asymptotic to the ray $\check{r}_0 :=\{y \vert \arg(y) = \theta - \frac{\pi}{d_2+1}$  and returns to $\infty$  in sector $\hat{S}_k$ asymptotically along the ray $\check{r}_k := \{y \vert \arg(y) = \theta + \frac{2(k-\frac{1}{2})}{d_2+1}$. These will  be called the ``anti-wedge'' contours.  (See Fig. 1.)    The choice of these contours is determined by the requirement that all moment integrals of the form
     \be
     \int_{\GH_k}  y^j e^{-V_2(y) +xy }  dy , \quad  \int_{\GC_k }y^k e^{V_2(y) -xy} dy,
     \quad k=1, \dots d_2, \quad j\in \Nbb
     \ee
    be uniformly convergent in $x\in \Cbb$.
     In the case when the other potential $V_1(x)$ is also a polynomial, of degree $d_1 +1$, the contours $\{\G_k\}_{k=1, \dots d_1}$ in the $x$--plane may be defined similarly.

The ``partition function'' is defined here to be the multiple integral \
\be
\mathcal Z_N:=\frac 1{N!} \iint_{\varkappa^N} \prod_{j=1}^{N} \d x_j \d y_j \Delta(X) \Delta(Y) \prod_{j=1}^N {\rm e}^{-V_1(x_j) - V_2(y_j) + x_jy_j}
\ee
where $\Delta(X)$ and $\Delta(Y)$ denote the usual Vandermonde determinants and the factor $\frac 1{N!}$ is chosen for convenience.

Such multiple integral can also be represented as the following determinant \
\be
\mathcal Z_N = \det[\mu_{ij}]_{0\leq i,j\leq N-1}\ ,\ \ \mu_{ij} := \int_{\varkappa} x^i y^j {\rm e}^{-V_1(x) - V_2(y) + xy}\d x \d y
\ee
The denomination of ``partition function'' comes from the fact \cite{MS, EM, BEH} that  when $\varkappa$ coincides with $\mathbb R\times \mathbb R$ then  $\mathcal Z_N$ coincides (up to a normalization for the volume of the unitary group) with the following matrix integral 
\be
\iint \d M_1 \d M_2 {\rm e}^{-\tr (V_1(M_1) + V_2(M_2) - M_1 M_2)}
\ee
extended over the space of Hermitean matrices $M_1, M_2$ of size $N\times N$, namely the normalization factor for the measure $\d \mu(M_1,M_2)$ introduced in \ref{101}.

   \begin{figure}
   \begin{center}
\includegraphics[width=.5\textwidth]{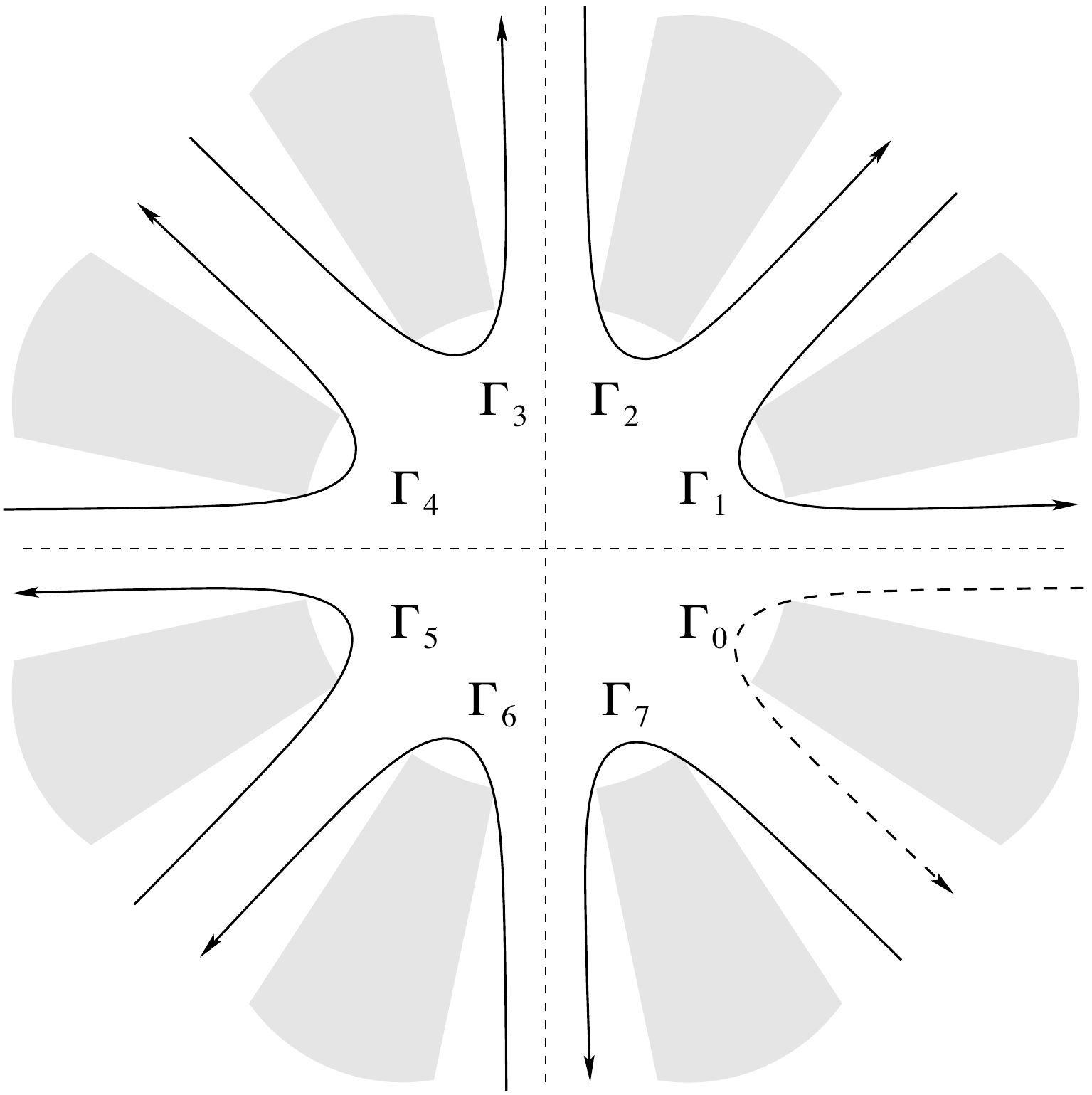}
\caption{ Wedge and anti-wedge contours for $V_2(y)$ of degree $D_2 +1$}
  \end{center}
  \end{figure}

\subsection{Riemann--Hilbert characterization for the orthogonal polynomials}
A Riemann--Hilbert characterization of the biorthogonal polynomials is a crucial step towards implementing a steepest--descent analysis. In our context it is also crucial in order to tie the random matrix side to the theory of isomonodromic deformations. 

We first recall the approach given by Kuijlaars and McLaughin (referred to as KM in the rest of the article) in \cite{KM}, suitably extended and adapted (in a rather trivial way) to the setting and notation of the present work. 
We quote -paraphrasing and with a minor generalization- their theorem, without proof.
\begin{theorem} [Kuijlaars and McLaughin asymptotic]
\label{KMthm}
The monic bi-orthogonal polynomial $\pi_n(x)$ is the $(1,1)$ entry of the solution $\Gamma(x)$ (if it exists) of the following Riemann-Hilbert problem for $\Gamma(x)$.
\begin{enumerate}
\item The matrix $\Gamma(x)$ is piecewise analytic in $\mathbb C\setminus \bigsqcup \Gamma_j$;
\item the (non-tangential) boundary values of $\Gamma(x)$ satisfy the relations 
\bea
\Gamma(x)_+ = \Gamma(x)_- \le[
\begin{array}{cccc}
 1& w_{j,1}&\dots & w_{j,d_2}\\
  &1&0&0\\
  &&\ddots&\\
  &&&1
\end{array}
\ri]\ ,\ \ \ x\in \Gamma_j\\
w_{j,\nu}= w_{j,\nu}(x):= {\rm e}^{-V_1(x)}  \sum_{k=1}^{d_2} \varkappa_{jk}\int_{\GH_k} y^{\nu-1} {\rm e}^{-V_2(y) +xy} \d y 
\eea
\item as $x\to \infty$ we have the following asymptotic expansion 
\be
\Gamma(x) \sim \left(I_d + \frac {Y_{N,1}}x +\mathcal{O}\left(\frac{1}{x^2}\right)\right)
\begin{pmatrix}
x^{N} & 0 &  0\\
0 & x^{-m_N-1}Id_{r_N}& 0\\
0&0& x^{-m_N}Id_{d_2-r_N} \\
\end{pmatrix}
\label{Gammaexpinf}
\ee
where we have defined the integers $m_N, r_N$ as follows
    \be 
    N = m_N d_2 + r_N, \quad m_N, r_N \in \mathbb{N}, \quad 0 \leq r_n \leq d_2-1
    \ee
\end{enumerate}
\end{theorem}

It follows from \cite{KM} that the solution $\Gamma_N(x)$ has the following form 
\bea
&&\Gamma_N(x):= \Gamma(x):= \le[
\begin{array}{cccc}
\pi_{N}(x) & \CC_0(\pi_{N}) & \dots &\CC_{d_2-1} (\pi_{N})\\
p_{N-1}(x) & \CC_0(p_{N-1})&\dots & \CC_{d_2-1}(p_{N-1})\\
\vdots &&&\vdots\\
p_{N-d_2}(x) & \CC_0(p_{N-d_2}) & \dots & \CC_{d_2-1} (p_{N-d_2})
\end{array}
\ri]\label{GammaCauchy}\ ,\\
&& \CC_i(f(z)):=\frac{1}{2\pi i}\int\!\!\!\int_\varkappa\frac{f(x)}{x-z}y^i\, e^{-V_1(x)-V_2(y)+xy}dydx
\eea
where the polynomials denoted above by $p_{N-1},\dots,p_{N-d_2}$ are some polynomials of degree not exceeding $N-1$, whose detailed properties are largely irrelevant for our discussion; we refer to \cite{KM} for these details.

By a left multiplication of this solution by a suitable constant matrix we can see that the matrix
\be
\widehat \Gamma_N:= \le[
\begin{array}{cccc}
\pi_{n} & \CC_0(\pi_{n}) & \dots &\CC_{d_2-1} (\pi_{n})\\
\pi_{n-1} & \CC_0(\pi_{n-1})&\dots & \CC_{d_2-1}(\pi_{n-1})\\
\vdots &&&\vdots\\
\pi_{n-d_2} & \CC_0(\pi_{n-d_2}) & \dots & \CC_{d_2-1} (\pi_{n-d_2})
\end{array}
\ri]
\ee
and $\Gamma_N$  are related as 
\beq
 \widehat \Gamma_N(x)=U_N \Gamma_N(x)
 \eeq
where $U_N$ is a constant matrix (depending on $N$ and on the coefficients of the polynomials but not on $x$).
As an immediate consequence, $\widehat \Gamma_N$ solves the same RHP as $\Gamma$ with the exception of the normalization at infinity (\ref{Gammaexpinf}).

The present RHP is not immediately suitable to make the connection to the theory of isomonodromic deformations as described in \cite{JMI, JMII}; we recall that this is the theory that describes the deformations of an ODE in the complex plane which leave  the Stokes' matrices  (i.e. the so--called  {\em extended monodromy data}) invariant.  The solution $\Gamma_N$ (or $\widehat \Gamma_N$) does not solve any ODE as formulated, because the jumps on the contours are non constant. If -however- we can relate $\Gamma_N$ with some other RHP with constant jumps, then its solution can be immediately shown to satisfy a polynomial ODE, which allows us to use the machinery of \cite{JMI,JMII}. This is the purpose of the next section.

\subsection{A RHP with constant jumps}

In \cite{BEH} the biorthogonal polynomials were characterised in terms of an ODE or --which is the same-- of a RHP with constant jumps. In order to connect the two formulations we will use some results contained in \cite{BHI} and we start by defining some auxiliary quantities:
for $1\leq k \leq d_2$, define the $d_2$ sequences of functions $\{\psi_m^{(k)}(x)\}_{m\in \mathbb{N}}$  as follows:
\be
 \psi_m^{(k)}(x) := \frac{1}{2\pi i}\int_{\GC_k}ds \int \!\!\!\int_{\varkappa}  dz dw 
\frac{\pi_m(z)e^{-V_1(z)}}{x-z}\frac{V_2'(s)-V_2'(w)}{s-w}  e^{-V_2(w)+V_2(s) +zw - xs }, \quad 1\leq k \leq d_2,
\label{psikmdef}
\ee
and let
\be
\psi^{(0)}_m(x):= \pi_m(x)e^{-V_1(x)}.
\ee

In terms of these define, for $N \ge d_2$, the sequence of $(d_2 + 1) \times (d_2 +1)$
matrix valued functions $\widehat{\PsiN}(x)$ 
\be
\widehat\PsiN(x):= \le[
\begin{array}{ccc}
\psi_N^{(0)}(x) & \dots& \psi_N^{(d_2)}(x)\\
\vdots &&\vdots\\
\psi_{N-d_2}^{(0)}(x) & \dots& \psi_{N-d_2}^{(d_2)}(x)
\end{array}\ri]
\label{hatpsi}
\ee
The following theorem is easily established using the properties of the bilinear concomitant and it is a very special case of the setting of \cite{Bertobisemi} (Cf. Appendix \ref{B} for a self-contained re-derivation)

\bt[Jump discontinuities in $\widehat{\PsiN}$]
The limits $\widehat{\PsiN}{}_{\pm}$ when approaching the
contours $\G_j$ from the left ($+$) and right($-$) are related by the following
jump discontinuity conditions
\bea
{\widehat \PsiN}{}_{+}(x) &\&= {\widehat \PsiN}{}_{-} (x)\Hb^{(j)}
\label{jumpdiscPsi} \\
\eea
where
\bea
\Hb^{(j)} &\& :=  \Ib  - 2\pi i \eb_0  \kappab^T \cr
\HbT{}^{\!\!(j)}&\&=  (\Hb^{(j)} )^{-1} =  \Ib  + 2\pi i \eb_0  \kappab^T
\label{HHT} \\
 \eb_0&\& := \begin{pmatrix}1 \cr 0 \cr \vdots \cr  0 \end{pmatrix}
\quad \kappab:= \begin{pmatrix} 0 \cr \varkappa_{j 1} \cr \vdots \cr \varkappa_{j d_2}\end{pmatrix}
\eea
\et
{The proof of this theorem is given in Appendix \ref{B}.}
For later convenience we define also 
\be
\PsiN:= U_N^{-1} \widehat\PsiN
\ee
The relationship with the matrices $\Gamma_N$,  $\widehat \Gamma_N$ introduced in the previous section  is detailed in the following 
\begin{theorem} [Factorization theorem]
\label{factorization}
\label{thmadditional}
The following identities hold
\be
{\widehat \PsiN}(x) = \widehat\Gamma_N(x)  V(x) W (x)\ ,\ \ 
\PsiN(x) = \Gamma_N(x) V(x) W(x)  
\ee
where
\be
V := \begin{pmatrix} {\rm e}^{-V_1(x)} & 0 \cr
                   0  &  V_0 ,\quad \end{pmatrix}\ ,\qquad  
W(x) :=  \begin{pmatrix} 1 & 0 \cr
                      0 &  W_0(x)\end{pmatrix}                    
\ee
 and $V_0$, $W_0(x)$ are the $d_2 \times d_2$ matrices with elements

\bea
(V_0)_{jk} &\&= 
\left[
\begin{array}{ccccc}
v_2 &v_3&\dots& & v_{d_2+1}\\
v_3& &&v_{d_2+1}&\\
&&\cdot^{\,\,\ds \cdot^{\,\,\ds \cdot}} &&\\
v_{d_2}&v_{d_2+1} &&&\\
v_{d_2+1} &&&&
\end{array}\right]
= \\
&\&= \begin{matrix} v_{j+k}  & {\rm if} \  j+k \leq d_2 +1\cr 
                                          0 &  {\rm if} \ j+k > d_2 +1, 
                                    \label{V0def}\end{matrix} \\ 
  (W_0(x))_{jk} &\& =\int_{\GC_k} y^{j-1} e^{V_2(y) - xy}  dy, \quad 1\leq j,k \leq d_2   
   \label{W0def}                              
\eea
  \end{theorem}

The proof is a direct verification by multiplication by matrices, noticing that the matrix  $V_0$ is nothing but the matrix representation of $\frac {V_2'(y)-V_2'(s)}{y-s}$ as a quadratic form in the bases $1,y, y^2,\dots, y^{d_2-1}$ and $1,s,s^2, \dots, s^{d_2-1}$ (more details are to be found on appendix \ref{A}, based on \cite{BHI, Bertobisemi})
The RHP for $\PsiN$ can be read off from that of $\Gamma_N$ and the fact that the jumps are constants. For convenience we collect the information in the following
\begin{theorem} 
\label{24}
The matrix $\PsiN$ is the unique solution of the following RHP:
\begin{enumerate} 
\item Constant Jumps: \bea
\PsiN{}_{+}(x) &\&= \PsiN{}_{-} (x)\Hb^{(j)} \\
\eea
\item Asymptotic at infinity:
\beq \PsiN(x)\sim\Gamma_N \begin{pmatrix}
x^{N}e^{-V_1(x)} & 0 &  0\\
0 & x^{-m_N-1}Id_{r_N}& 0\\
0&0& x^{-m_N}Id_{d_2-r_N} \\
\end{pmatrix}  \Psi_0(x) 
\eeq
where 
\beq {\Gamma_N}=Id+\frac{{Y}_{N,1}}{x}+...\eeq 
and where $\Psi_0(x):=V(x)W(x)$ will be referred to as  the {\em bare} solution. Its asymptotic at infinity can be computed by steepest descent, but since it is $N$--independent, for the sake of brevity, we do not report on it (details are contained in \cite{BEH2, BHI}).
\item $\Psi_N$ has constant jumps 
\item $\Psi_{N}'(x)\Psi_N^{-1}=D_N(x)$ where $D_N(x)$ is a polynomial in $x$
\item $\partial_{u_K}\Psi_{N}(x)\Psi_N^{-1}=U_{K,N}(x)$ is polynomial in $x$.
\item $\partial_{v_J}\Psi_{N}(x)\Psi_N^{-1}=V_{J,N}(x)$ is polynomial in $x$.
\item $\det(\Psi_{N+1}\Psi_N^{-1})=Cste$
\end{enumerate}
\end{theorem}
The points (4,5,6,7) in the above theorem can be found in \cite{BE, BEH2}

In the next section we shall define a proper notion of isomonodromic tau function: it should be pointed out that the definition of \cite{JMI,JMII} cannot be applied as such because --as showed in \cite{BEH2}-- the ODE that the matrix $\PsiN$ (or $\widehat \PsiN$) solves, has a highly degenerate leading coefficient at the singularity at infinity.


%

In the list, the crucial ingredients are the differential equations (in $x$ or relatively to the parameters $u_K$ and $v_J$). First, the fact that $D_N(x)$ is a polynomial comes from explicit computation (See \cite{BE} for example). The result concerning the determinant of $R_N(x)$ can also be found in \cite{BE} where one has: $\det(\Psi_{N+1}\Psi_N^{-1})=\det(a_N(x))=Cste$.  The properties concerning the differential equations relatively to parameters can be found in \cite{BE} too.
Under all these assumptions, we will show that the proof of Jimbo-Miwa-Ueno can be adapted and that we can define a suitable $\tau$-function in the same way Jimbo-Miwa-Ueno did it.

\section{Definition of the $\tau$-function}
\label{taudef}
In this section, we will place ourselves in a more general context than the one described above; we will show that under few assumptions one can define a good notion of  tau-function. 

 More generally we will denote with $t_a$ the isomonodromic parameters (in our case they are the $u_K$'s and ths $v_J$'s) and a subscript $a$ or $b$ is understood as a derivation relatively to $t_a$ or $t_b$. 
For a function $f$ of the isomonodromic times we will denote by the usual symbol its differential
 \beq
  \d f= \sum_a \partial_{t_a}f \d t_a=\sum_a f_a 
  \d t_a
 \eeq
Our setup falls in the following framework that it is useful to ascertain from the specifics of the case at hands.
Suppose we are given a matrix 
\beq
 \Psi(x) \sim  Y(x)\,  \Xi(x)\ ,
 \ \ Y(x):= \le(\mathbf 1 + \frac{Y_1}x + \frac {Y_2}{x^2}+ \dots \ri) x^S
 \eeq 
 where $\Xi(x) = \Xi(x; \mathbf t)$ is some explicit expression (the ``bare'' isomonodromic solution) and $S$ is a matrix independent of the isomonodromic times. This implies that if we define 
 the one--form-valued matrix $\H(x;\mathbf t)$ by 
  \be
 \H(x; \mathbf t) = \d \Xi(x; \mathbf t)\, \Xi(x;\mathbf t)^{-1}
 \ee
then $\H(x) = \sum \H_a \d t_a$ (we suppress explicit mention of the $\mathbf t$ dependence henceforth) is some solution of the zero-curvature equations:
\be
\pa_a \H_b - \pa _b \H_a = [\H_a,\H_b]
\label{barezcc}
\ee
We will {\bf assume} (which is the case in our setting) that all $\H_a$ are {\bf polynomials} in $x$. We will also use that the dressed deformations $\Omega_a$ given by $\Psi_a=\Omega_a \Psi$ are polynomials. Moreover, according to the asymptotic they are given by: 
\be
\Omega _a = (Y \H_a Y^{-1})_{pol}.
\ee
In this very general (and generic) setting we can formulate the definition of a ``tau function'' as follows
\begin{definition}
The tau-differential is the one-form defined by 
\be
\omega:= \sum \omega_a \d t^a:=\sum_a \res{} \tr \le(Y^{-1}Y' \H_a \ri) \d t^a
\ee
\end{definition}
The main point of the matter is that -without any further detail- we can now prove that the tau-differential is in fact closed and hence locally  defines a function.
\bt
\label{closurethm}
The tau-differential is a closed differential and locally defines a $\tau$--function as 
\be
\d \log \tau = \omega
\ee 
\et
{\bf Proof.} 
We need to prove the closure of the differential.
We first recall the main relations between the bare and dressed deformations
\be
\pa_a Y = \Omega_a Y - Y \H_a\  \ ; \qquad
Y\H_a Y^{-1} =  \Omega_a - \mathcal R_a\ ;\ \qquad \mathcal R_a:= \pa_a Y Y^{-1}\label{defs}
\ee
We note that -by construction- $\Omega_a = (Y\H_aY^{-1})_{pol}$ is a polynomial while $\mathcal R_a= \mathcal O(x^{-1})$ {\em irrespectively of the form of $S$}.
We compute the cross derivatives directly 
\bea
\pa_a \omega_b &=&
\restr\bigg(
-Y^{-1} \le(\Omega_a Y - Y \H_a\ri) Y^{-1} Y' \H_b + Y^{-1} \le( \Omega_a Y - Y \H_a \ri)' \H_b +  Y^{-1} Y' \pa_a \H_b
\bigg) \cr
 &=& \restr \bigg(
\H_a Y^{-1} Y' \H_b + Y^{-1} \Omega_a' Y \H_b  - Y^{-1} Y'  \H_a \H_b - \H_a' \H_b + Y^{-1} Y' \pa_a H_b
\bigg) \cr
&=& \restr \bigg(
Y^{-1} Y' \le([\H_b, \H_a] + \pa_a \H_b\ri) + Y^{-1} \Omega_a' Y \H_b - \overbrace{\H_a' \H_b}^{\hbox{polynomial}}\bigg)\cr
&=& \restr \bigg(
Y^{-1} Y' \le([\H_b, \H_a] + \pa_a \H_b\ri) - \Omega_a' \mathcal R_b   
\bigg)
\eea
where, in the last step, we have used that $Y  \H_b Y^{-1} = \Omega_b - \mathcal R_b$ and that the contribution coming from $\Omega_b$ vanishes since it is a polynomial.
Rewriting the same with $a\leftrightarrow b$ and subtracting we obtain 
\bea
&\& \pa_a \omega_b - \pa_b \omega_a = \restr \bigg(
2 Y^{-1} Y' [\H_b, \H_a] - \Omega_a' \mathcal R_b + \Omega_b' \mathcal R_a  + Y^{-1}Y' \le(\pa_a \H_b - \pa_b \H _a\ri)
\bigg) \cr 
&\&=  \restr \bigg(
 Y^{-1} Y' [\H_b, \H_a] - \Omega_a' \mathcal R_b + \Omega_b' \mathcal R_a  + Y^{-1}Y' \big(\overbrace{\pa_a \H_b - \pa_b \H _a + [\H_b, \H_a]}^{=0\hbox { by the ZCC \ref{barezcc}}}\big)
\bigg)\cr
&\& = \restr \bigg(
 Y^{-1} Y' [\H_b, \H_a] - \Omega_a' \mathcal R_b + \Omega_b' \mathcal R_a  \bigg) 
 \label{closure}
\eea
Note that, up to this point, we only used the zero curvature equations for the connection $\nabla = \sum(\pa_a - \H_a )\d t^a$ and the fact that $\H_a$ are polynomials in $x$. 
We thus need to prove that the last quantity in (\ref{closure}) vanishes: this follows from the following computation, which uses once more the fact that $\H_a$ and $\Omega_a$ are all polynomials. Indeed, we have $\restr (\H_a' \H_b) = 0$ and hence (using (\ref{defs}))
\bea
0&\& = \res{}\tr (\H_a' \H_b)=\restr\le( \le(Y \H_a Y^{-1}\ri)' Y \H_b Y^{-1}\ri) - \restr \le(Y' \H_a \H_b Y^{-1}\ri) +  \restr\le(
\H_a Y^{-1} Y' \H_b
\ri) \cr
&\& = \restr \le(\le(\Omega_a - \mathcal R_a\ri)' \le(\Omega_b - \mathcal R_b\ri)\ri)  + 
\restr \le(
Y^{-1} Y' \le[\H_b, \H_a\ri]\ri)  \cr
\&\&=
\restr \bigg( 
\overbrace{\Omega_a'\Omega_b}^{\hbox{poly}}  - \mathcal R_a' \Omega_b - \Omega_a' \mathcal R_b +\overbrace{ \mathcal R_a' \mathcal R_b}^{=\mathcal O(x^{-2})} + Y^{-1}Y' [\H_b,\H_a]
\bigg)
\cr
&\&  = \restr \bigg( 
 - \mathcal R_a' \Omega_b - \Omega_a' \mathcal R_b  + Y^{-1}Y' [\H_b,\H_a]
\bigg)=0
\eea 
Using integration by parts (and cyclicity of the trace) on the first term here above, we obtain precisely the last quantity in (\ref{closure}). The Theorem is proved. {\bf Q.E.D.}

\subsection{Application to our problem}
\label{application}
We now apply the general definition above to our setting, with the identifications $\Psi = \Psi_N$, $Y=\Gamma_N$ (as a formal power series at $\infty$) and  $\Xi = \Psi_0$. 
We will write $Y_N$ instead of $\Gamma_N$ in the expressions below to emphasize that we consider its asymptotic expansion at $\infty$
 This reduces the definition of the tau function to the one below
\begin{definition}
\label{tauN}
The $\tau$-function is defined by the following PDE 
\beq  d (\log \tau_N) =\Res_{x \to \infty} \Tr\left(Y_N^{-1}Y_N'  \d(\Psi_0)\Psi_0^{-1}\right) \eeq
where $Y_N$ is the formal  asymptotic expansion of $\Gamma_N$ at infinity
\beq Y_N=\widetilde Y _N\begin{pmatrix} 
x^{N} & 0 &  0\\
0 & x^{-m_N-1}Id_{r_N}& 0\\
0&0& x^{-m_N}Id_{d_2-r_N} \\
\end{pmatrix}\eeq
\end{definition}

\begin{remark}
The matrix $S$ of the previous section  in our case becomes:
\beq S=\begin{pmatrix} 
N & 0 &  0\\
0 & (-m_N-1)\, Id_{r_N}& 0\\
0&0& -m_N\, Id_{d_2-r_N} \\
\end{pmatrix}\eeq
The partial derivatives of $\ln \tau_N$  split into two sets which have different form:
\beq \partial_{u_K} \log \tau_N=-\Res_{x \to \infty} \Tr\left(Y_N^{-1}Y_N' \frac{x^K}{K} \bf{E}_{11} \right)\eeq
\bea 
\partial_{v_J} \log \tau_N&=&\Res_{x \to \infty}\Tr\left(Y_N^{-1}Y_N'  \partial_{v_J}(\Psi_0)\Psi_0^{-1}\right)\eea
where in the last equation the term $\partial_{v_J}(\Psi_0)\Psi_0^{-1}$ has non-zero entries only in the anti-principal minor of size $d_2$.
\end{remark}

One can notice that the situation we are looking at is a generalization of what happen in the one-matrix case. In the 1-matrix model, the matrix $S$ is zero and therefore $Y_N$ are (formal) Laurent series. The matrix $\Psi_0$ matrix is absent in that case since there is only one potential and thus one recovers the usual definition of isomonodromic tau function (see \cite{BEH4}). Note also that in the derivation with respect to $v_J$ we have obtained the second equality using  the block diagonal structure of $\Psi_0$ (first row/column does not play a role). It is remarkable that the two systems are completely decoupled, i.e. that in the first one the matrix $\Psi_0$ (containing all the dependance in $V_2$) disappears and that in the second one the matrix $A_0$ (containing the potential $V_1$) also disappears. 

\subsection{Discrete Schlesinger transformation: Tau-function quotient}  
\label{Schlesinger}
In this section we investigate the relationship between the tau-function of Def. \ref{tauN} and the partition function $\mathcal Z_N$ of the matrix model.  

We anticipate that the two object turn out to be the same (up to a nonzero factor that will be explicitly computed, Thm. \ref{main}): the proof relies on two steps, the first of which we prepare in this section. 
These are
\begin{itemize}
\item proving that they satisy the same recurrence relation
\item identifying the initial conditions for the recurrence relation.
\end{itemize}

We start by investigating the relationship between $\tau_N$ and $\tau_{N+1}$; this analysis is essentially identical to the theory developed in \cite{JMII} and used in \cite{BE_mom}, but we report it here for the convenience of the reader.

From the fact that the $\Psi_N$ has constant jumps, we deduce that $\Psi_{N+1}\Psi_N^{-1}$ is an entire function. Moreover asymptotically  it looks like:
\bea \Psi_{N+1}\Psi_N^{-1}&=&\tilde{Y}_{N+1}\begin{pmatrix}
x^{N+1}e^{-V_1(x)} & 0 &  0\\
0 & x^{-m_{N+1}-1}Id_{r_{N+1}}& 0\\
0&0& x^{-m_{N+1}}Id_{d_2-r_{N+1}} \\
\end{pmatrix}  \Psi_0(x) \cr & & \Psi_0(x)^{-1} \begin{pmatrix}
x^{-N}e^{V_1(x)} & 0 &  0\\
0 & x^{m_N+1}Id_{r_N}& 0\\
0&0& x^{m_N}Id_{d_2-r_N} \\
\end{pmatrix}\tilde{Y}_N\eea
\beq \Psi_{N+1}\Psi_N^{-1}=\tilde{Y}_{N+1}\begin{pmatrix}
x & 0 &  0&0\\
0 &Id_{r_N-1} &0&0\\
0& 0&x^{-1}& 0\\
0&0& 0&Id_{d_2-1-r_N} \\
\end{pmatrix}\tilde{Y}_N\eeq
Thus, remembering that $\tilde{Y}_N$ is a series $x^{-1}$, Liouville's theorem states that $\Psi_{N+1}\Psi_N^{-1}$ is a polynomial of degree one, and hence, for some constant matrices $R_N^0, R_N^1$ we must have
\beq
 \Psi_{N+1}\Psi_N^{-1}=R_N(x)=R_N^0+x R_N^1
 \eeq

From the fact that $\det(R_N)$ does not depend on $x$ (last property Thm. \ref{thmadditional}), we know that $R_N^{-1}(x)$ is a  polynomial of degree at most one as well (this is easy if one consider the expression of the inverse of a matrix using the co-matrix).

Comparing the asymptotics of $\Psi_{N+1}$ and $R_N(x) \Psi_N$ term-by-term in the expansion in inverse powers of $x$ and after some elementary algebra one obtains (\cite{JMI} Appendix A):
\beq
 R_N(x)=E_{\alpha_0}x+R_{N,0} \qquad \hbox{and} \qquad R_N^{-1}(x)=E_1 x+R_{N,0}^{-1} 
 \eeq
Here we have introduced the notation $\alpha_0=r_N+1$ which corresponds to the index of the column where the coefficient $x^{-1}$ is to be found in the asymptotic of ${\Psi}_{N+1}{\Psi}_N^{-1}$. This notation is the standard notation used originally by Jimbo-Miwa in a Schlesinger transformation.
The matrix $(R_{N,0})_{\alpha,\beta}$ is given by:
\beq \begin{array}{cccc}
&\beta=\alpha_0&\beta=1 & \beta \neq \alpha_0,1\\
\\
\alpha=\alpha_0 & \frac{-(Y_{N,2})_{\alpha_0,1}+\sum_{\gamma \neq \alpha_0}(Y_{N,1})_{\alpha_0,\gamma}(Y_{N,1})_{\gamma,1}}{(Y_{N,1})_{\alpha_0,1}} & -(Y_{N,1})_{\alpha_0,1}& -(Y_{N,1})_{\alpha_0,\beta} \\
\\
\alpha=1 & \frac{1}{(Y_{N,1})_{\alpha_0,1}} & 0 &0 \\
\\
\alpha \neq \alpha_0, 1 & -\frac{(Y_{N,1})_{\alpha,1}}{(Y_{N,1})_{\alpha_0,1}} & 0 & \delta_{\alpha, \beta} \\
\end{array} \eeq

and $(R_{N,0}^{-1})_{\alpha,\beta}$ is given by:
\beq \begin{array}{cccc}
&\beta=\alpha_0&\beta=1 & \beta \neq \alpha_0,1\\
\\
\alpha=\alpha_0& 0 & (Y_{N,1})_{\alpha_0,1}& 0\\
\\
\alpha=1 & -\frac{1}{(Y_{N,1})_{\alpha_0,1}} & -\frac{-(Y_{N,2})_{\alpha_0,1}}{(Y_{N,1})_{\alpha_0,1}}+(Y_{N,1})_{1,1} & -\frac{(Y_{N,1})_{\alpha_0,\beta}}{(Y_{N,1})_{\alpha_0,1}} \\
\\
\alpha \neq \alpha_0, 1 & 0& (Y_{N,1})_{\alpha,1} & \delta_{\alpha, \beta}\\
\end{array} \eeq

While the formulae above might seem complicated, we will use  the two important observations:
\beq E_{\alpha_0}R_{N,0}^{-1}+R_{N,0}E_1=R_{N,0}^{-1}E_{\alpha_0}+E_1R_{N,0}=0\eeq
\begin{center} $ R_N^{-1}(x)R_N'(x)=R_{N,0}^{-1}E_{\alpha_0}$ does not depend on $x$. \end{center}

The recurrence relation satisfied by the sequence  $\{\tau_N\}$ is derived in the next theorem.
\begin{theorem} 
Up to multiplication by functions that do not depend on the isomonodromic parameters (i.e. independent of the potentials $V_1,V_2$) the following identity holds
\beq  
\frac{\tau_{N+1}}{\tau_N}=(Y_1)_{1,\alpha_0}
\eeq
\end{theorem}
{\bf Proof}
The proof follows \cite{JMII} but we report it here for convenience of the reader.
Consider the following identity
\beq \Psi_{N+1}=Y_{N+1}\Psi_0=R_NY_N\Psi_0\eeq
This implies that 
\beq
 Y_{N+1}=R_NY_N
 \eeq
Taking the derivative with respect to $x$ gives:
\beq Y_{N+1}^{-1}Y_{N+1}'=Y_N^{-1}R_N^{-1}R_N'Y_N+Y_N^{-1}Y_N'\eeq
Therefore we have:
\bea 
\d \log \tau_{N+1}-\d \log \tau_N &=&\Res_{x \to \infty} \Tr( (Y_N^{-1}R_N^{-1}R_N'Y_N+Y_N^{-1}Y_N'-Y_N^{-1}Y_N) \d(\Psi_0 )\Psi_0^{-1}) \cr
&=&\Res_{x \to \infty} \Tr( Y_N^{-1}R_N^{-1}R_N'Y_N\d(\Psi_0 )\Psi_0^{-1}) 
\eea
We now need to ``transfer'' the exterior derivative from $\Psi_0$ to $Y_N$. This can be done using  that $\PsiN =Y_N \Psi_0$, so that
$$\d\PsiN=\d(Y_N) \Psi_0 + Y_N  \d(\Psi_0)$$
Equivalently:
\beq Y_N\d \Psi_0 \Psi_0^{-1}Y_N^{-1}=d(\PsiN)\PsiN^{-1}-dY_NY_N^{-1}\eeq
Inserting these identities  in the tau quotient we obtain the relation
\beq
 d \log \tau_{N+1}-d \log \tau_N =\Res_{x \to \infty} \Tr\le (R_N^{-1}R_N' d(\PsiN)\PsiN^{-1}-R_N^{-1}R_N'dY_NY_N^{-1}\ri)\eeq

The first term is residueless at $\infty$ since $\d \PsiN \PsiN^{-1}$ is polynomial in $x$ and $R_N^{-1}R_N'$ does not depend on $x$. Therefore we are left only with:
\beq d \log \tau_{N+1}-d \log \tau_N =-\Res_{x \to \infty} \Tr(R_N^{-1}R_N'dY_NY_N^{-1})\eeq
A direct matrix computation using the explicit form of $R_N$ yields
\beq 
\d \log \tau_{N+1}-\d \log \tau_N =\d \log((Y_{N,1})_{1,\alpha_0})
\eeq
and hence
\beq  
\frac{\tau_{N+1}}{\tau_N}=(Y_1)_{1,\alpha_0}
\eeq
The last equality is to be understood up to a multiplicative constant not depending on the parameters $u_K$ and $v_J$ in $\tau$. {\bf Q.E.D.}\par \vskip 5pt

In order to complete the first step we need to express the entry ${(Y_1)}_{1,\alpha_0}$  in terms of the ratio of two consecutive partition functions. This is accomplished in the following section.

\begin{theorem}
For the matrix $ \Gamma_N $ the asymptotic expansion at infinity (\ref{Gammaexpinf}) is such that 
\be 
(Y_{N,1})_{1,\alpha_0} = (v_{d_2+1})^ S h_N = (v_{d_2+1})^ S \frac {\mathcal Z_{N+1}}{\mathcal Z_N}
\ee
where $S$ and $\alpha_0\in \{0,1,\dots, d_2-1\}$ are defined by the following relation
\be
N = d_2 S + \alpha_0-1 
\ee
\end{theorem}
{\bf Proof}
In order to compute $(Y_{N,1})_{1,\alpha_0}$ it is sufficient to compute the leading term of the expansion at $\infty$ appearing in the first row of the matrix $\Gamma_N$. Recalling the expression (\ref{GammaCauchy}), we start by the following direct compuation using integration by parts
\bea 
\int\!\!\!\int_\kappa dzdw \,\pi_N(z)z^iw^{k-1}e^{-V_1(z)-V_2(z)+zw} 
\&\& =\int\!\!\!\int_\kappa dzdw \,\pi_N(z)e^{-V_1(z)}w^{k-1}e^{-V_2(w)}\frac{d^i}{dw^i}\left(e^{zw}\right) \cr
&\&=(-1)^i\int\!\!\!\int_\kappa dzdw \,\pi_N(z)e^{-V_1(z)+zw}\frac{d^i}{dw^i}\left(w^{k-1}e^{-V_2(w)}\right)\cr
&\&=\int\!\!\!\int_\kappa dzdw \, \pi_N(z) q_{d_2i+k-1}(w)e^{-V_1(z)-V_2(z)+zw}
\eea
where  $q_{d_2i+k-1}(w)$ is a polynomial of the indicated degree  whose leading coefficient is $v_{d_2+1}^i$. The last RHS is $0$ if $d_2i+k-1<N$ because of orthogonality. If $d_2i+k-1=N$ the integral gives $v_{d_2+1}^i h_N$ by the normality conditions concerning our biorthogonal set. 
This computation allows us to expand the Cauchy transform of $(\Gamma_{N})_{1,\alpha_0}$ near $\infty$ as follows:
\bea \mathcal{C}(p_Nw_{\alpha_0}(x))&=& \frac{1}{2\pi i}\int\!\!\!\int_\kappa dzdw \frac{\pi_N(z)}{z-w} w^{\alpha_0-1}e^{-V_1(z)-V_2(z)+zw} \cr
&=&-\sum_{i=0}^{S-1}\frac{1}{2\pi i}\int\!\!\!\int_\kappa dzdw \pi_N(z)\frac{z^i}{x^{i+1}} w^{\alpha_0-1}e^{-V_1(z)-V_2(z)+zw}\cr
&+& \frac{1}{2\pi i}\frac{1}{x^{S+1}}\int\!\!\!\int_\kappa dzdw \frac{\pi_N(z)}{x-z}z^{S} w^{\alpha_0-1}e^{-V_1(z)-V_2(z)+zw}+ \mathcal O(x^{-S-2})\cr
\eea
By orthogonality the first sum vanishes term-by-term and the leading coefficient of the second term is $v_{d_2+1}^S h_N$. {\bf Q.E.D.}\par\vskip 5pt

Recalling that the $\tau$-function is only defined up to a multiplicative constant not depending on $N$ nor on the coefficients $u_k$and $v_j$, we have
\beq \frac{\tau_{N+1}}{\tau_N}=(v_{d_2+1})^ {S_N} \frac{\mathcal Z_{N+1}}{\mathcal Z_N}\eeq
where $N = d_2 S_N + \alpha_0-1$
Hence for every $n_0$:
\beq \tau_{N}\mathcal Z_{n_0}=\mathcal Z_N \tau_{n_0}(v_{d_2+1})^{\sum_{j=n_0}^{N-1}S_j} \eeq
One would like to take $n_0=0$ because it enables explicit computations. As we will prove now there is a way of extending naturally all the reasoning down to $0$.

The RHP for $\Gamma_N$ (Thm. \ref{KMthm})  is perfectly well--defined for $N=0$ and has solution
\beq 
{\Gamma_0}=\begin{pmatrix}
1& \mathcal C_0(1)&\mathcal C_1(1)& \ldots &\mathcal C_{d_2-1}(1)\\
0&1&0&\ldots&0\\
\vdots&\ddots&\ddots&\ddots&0\\
\vdots&\ddots&\ddots&\ddots&0\\
0&0&\ldots&\ddots&1
\end{pmatrix}\ .
\eeq
Consequently we can take
\beq 
\tau_{N}\mathcal Z_0=(v_{d_2+1})^{\sum_{j=0}^{N-1}S_j} \mathcal Z_N \tau_0 
\eeq
Also note that $\mathcal Z_0\equiv 1$ (by definition).

We can compute $\tau_0$ directly from Def. \ref{tauN} because of the particularly simple and explicit expression of $\mathop{\Psi}_0= \Gamma_0 \Psi_0$.
\be
\d \ln \tau_0 = \restr\le(Y_0^{-1}Y_0' \d \Psi_0 \Psi^{-1}\ri) 
\ee
We claim that this expression is identically zero (and hence we can define $\tau_0\equiv 1$); indeed,  \beq 
Y_0^{-1}Y_0'=\begin{pmatrix}
0& *&\dots&*\\
0&0& \dots &0\\
\vdots&\vdots& \ddots&0\\
0&0&\dots&0\\
\end{pmatrix}
\eeq 
and 
\beq 
\d \Psi_0(x)\Psi_0^{-1}(x)=\begin{pmatrix}
\star& 0&\dots&0\\
0&\star& \dots &\star\\
\vdots&\vdots& \ddots&\vdots\\
0&\star&\dots&\star\\
\end{pmatrix}
\eeq 
so that the trace of the product is always zero (even before taking the residue). Combining the two results together gives the following theorem:

\begin{theorem} 
\label{main}
The isomodromic $\tau$-function and the partition function are related by:
$$
\forall N \in \mathbb{N}: \mathcal Z_N=(v_{d_2+1})^{\sum_{j=0}^{N-1}S_j} \tau_N \label{PartFuntAndIsoTau}
$$
where we recall that $S_j$ is given by the decomposition of $j+1$ in the Euclidian division by $d_2$: $S_j=E\left[\frac{j+1}{d_2}\right]$. A short computation of the power in $v_{d_2+1}$ gives:
$$
\forall N \in \mathbb{N}: \mathcal Z_N=(v_{d_2+1})^{d_2\frac{\alpha_N(\alpha_N-1)}{2}+\alpha_N(N-\alpha_N d_2)} \tau_N \label{PartFuntAndIsoTau2}$$
where $\alpha_N=E\left[\frac{N}{d_2}\right]$
\end{theorem}

The presence of the power in $v_{d_2+1}$ is due to a bad normalisation of the partition function itself ($\mathcal Z_N$) and can be easily cancelled out by taking $v_{d_2+1}=1$ from the start (it is just a normalization of the weight function). Moreover it is not surprising because in the work of Bergere and Eynard \cite{BgE}, all results concerning the partition function and its derivatives with respect to parameters have special cases for $u_{d_1+1}$ and $v_{d_2+1}$. It also signals the fact that the RHP is badly defined when $v_{d_2+1}=0$ because the contour integrals involved diverge and the whole setup breaks down. Indeed if $v_{d_2+1}=0$ this simply means that $V_{2}$ is a polynomial of lower degree and thus the RHP that we should set up should be of smaller size from the outset.

\section{Outlook}

In this article, we have restricted ourselves to contours going from infinity to infinity. This allows us to use integration by parts without picking up any boundary term. A natural extension of this work could be to see what happens when contours end in the complex plane, and especially study what happens when the end points moves (models with hard edges). This generalization is important in the computation of the gap probabilities of the Dyson model \cite{TWDyson}, which correspond to a random matrix model with Gaussian potentials but with the integration restricted to intervals of the real axis.

\section*{Acknowledgements}
We would like to thank John Harnad for proposing the problem, Seung Yeop Lee and Alexei Borodin for fruitful discussions. This work was done at the University of Montr\'eal at the departement of mathematics and statistics and the Centre de Recherche Math\'ematique (CRM) and O.M.  would like to thank both for their hospitality.
This work was partly supported
by the Enigma European network MRT-CT-2004-5652,
by the ANR project G\'eom\'etrie et int\'egrabilit\'e en physique 
math\'ematique  ANR-05-BLAN-0029-01,
by the Enrage European network MRTN-CT-2004-005616,
by the European Science Foundation through the Misgam program,
by the French and Japanese governments through PAI Sakurav,
by the Quebec government with the FQRNT.

\setcounter{section}{0}
\appendix
\renewcommand{\theequation}{\Alph{section}.\arabic{equation}}
\section{Factorization of ${\Psi_N}$}
\label{A}
Starting from the definition of  the last $d_2$ columns of $\widehat{\PsiN}$ (\ref{hatpsi})  we observe that
\bea
\psi_m^{(k)}(x) &\& := \frac 1{2i\pi} \int_{\GC_k} \!\!\!{\rm d}s \int\!\!\!\int_{\varkappa} \frac {\pi_m(z)}{x-z} \frac {V_2'(s)- V_2'(w)}{s-w} {\rm e}^{-V_1(z)-V_2(w)+V_2(s) + zw -xs}dwdz \\
&\&= \sum_{p,q}v_{q+p}  \frac 1{2i\pi}  
\int\!\!\!\int_{\varkappa} \frac {\pi_m(z)}{x-z} w^{p-1} {\rm e}^{-V_1(z)-V_2(w)+zw} \int_{\GC_k} {\rm d}s s^{q-1} {\rm e}^{V_2(s) -xs}\cr
&=&\sum_{p,q}(\widehat\Gamma_N)_{m,p}(V_0)_{p,q}(W_0)_{q,k}=(\widehat \Gamma_N \, V_0W_0)_{m,k}
\eea
This proves Thm. \ref{factorization}.

\section{Bilinear concomitant as intersection number}
\label{B}
We recall very briefly the result of \cite{Bertobisemi} stating that
\be
\frac{V_2'(\pa_x) - V_2'(-\pa_z)}{\pa_x+\pa_z} w(x)f(z)\bigg|_{z=x} =\int_\Gamma \int_{\check \Gamma} \frac {V_2'(\eta)-V_2'(s)}{\eta-s} {\rm e}^{x(\eta-s) -V_2(\eta)+V_2(s)} = 2i\pi \Gamma\# \check \Gamma =\hbox{constant} \ .
\ee
The last identity is obtained by integration by parts and shows that
the bilinear concomitant is just the intersection number of the
(homology classes) of the contours $\Gamma, \check \Gamma$. 
More precisely we get that:
\bea &&\frac{d}{d x}\int_\Gamma \int_{\check \Gamma}ds d\eta \, \frac {V_2'(\eta)-V_2'(s)}{\eta-s} {\rm e}^{x(\eta-s) -V_2(\eta)+V_2(s)} \cr
&=&\int_\Gamma \int_{\check \Gamma}ds d\eta \, (V_2'(\eta)-V_2'(s)) {\rm e}^{x(\eta-s) -V_2(\eta)+V_2(s)} \cr
&=&\int_\Gamma \int_{\check \Gamma}ds d\eta \, \frac{\partial}{\partial \eta}(-e^{-V_2(\eta)})e^{x\eta}e^{-xs+V_2(s)} -\int_\Gamma \int_{\check \Gamma}d\eta ds \, \frac{\partial}{\partial s}(e^{V_2(s)})e^{-xs}e^{x\eta -V_2(\eta)} \cr
&=&x\int_\Gamma \int_{\check \Gamma}ds d\eta \, e^{x\eta-xs-V_2(\eta)+V_2(s)}-x\int_\Gamma \int_{\check \Gamma}ds d\eta e^{x\eta-xs-V_2(\eta)+V_2(s)}\cr
&=&0
\eea
The matrix expression shows that the pairing is indeed a duality since the determinant is nonzero.
The undressing matrix $\Psi_0$ (that was originally introduced in Thm. \ref{24})  is thus
\be
\Psi_0 =\le[
\begin{array}{c|c}
1& \\
\hline
& \begin{array}{ccccc}
v_2 &v_3&\dots& & v_{d_2+1}\\
v_3& &&v_{d_2+1}&\\
&&\cdot^{\,\,\ds \cdot^{\,\,\ds \cdot}} &&\\
v_{d_2}&v_{d_2+1} &&&\\
v_{d_2+1} &&&&
\end{array}
\end{array}\ri]
\le[ \begin{array}{c|cccc}
1 &&&&\\
\hline
&f_1&f_2 &\dots &f_{d_2}\\
&f_1'&f_2'&\dots &f_{d_2}'\\
 & \vdots &&&\vdots\\
 & f_1^{(d_2-1)}&\dots && f_{d_2}^{(d_2-1)}
\end{array}\ri]
\ee
where the Wronskian subblock in the second term is constructed by
choosing $d_2$ homologically independent contour classes for the
integrations $\check \Gamma$;
\be
f_k(x):=\int_{\check \Gamma_k} {\rm e}^{-xs+V_2(s)}{\rm d}s\ ,\ \
k=1,\dots,d_2\ .
\ee
The dressing matrix $\Psi_0$ exhibits a 
  Stokes' phenomenon (of Airy's type) which is the inevitable drawback of removing
the $x$-dependence from the jump matrix.
We can now compute the jumps and see that it does not depend on $x$. For the  $k$-th column we have:
\be \psi_m^{(k)}(x) := \frac{1}{2\pi i} \int_{\GC_k}ds \int \!\!\!\int_{\varkappa}  dz dw 
\frac{\pi_m(z)e^{-V_1(z)}}{x-z} \frac{V_2'(s)-V_2'(w)}{s-w}  e^{-V_2(w)+V_2(s) +zw - xs }, \quad 1\leq k \leq d_2\ee
gives:
\bea
\psi_m^{(k)}(x)_+ &=& \psi_m^{(k)}(x)_-+  \psi_m^{(0)}(x)\int\int dsdw 
\frac{V_2'(s)-V_2'(w)}{s-w}  e^{-V_2(w)+V_2(s) +x(w - s) }
\\ 
&=&\psi_m^{(k)}(x)_- + \psi_m^{(0)}(x)\sum_{j=1}^{d_2}\varkappa_{\ell j}( \Gamma^{(y)}_j\# \check \Gamma_k)\ ,
\eea 


\end{document}